\documentclass{article}
\usepackage[T2A]{fontenc}
\usepackage[russian]{babel}
\usepackage{graphicx}
\usepackage{amsbsy}
\usepackage{amssymb}
\begin{document}
\makeatletter \Large \baselineskip 6mm \centerline{\bf On the Dependence of the Lifetime of an Atomic Cluster}
\centerline{\bf on the Intensity of Its Heat Exchange with the Environment}

\vskip 4mm

\centerline{A.I. Podlivaev$^{1)}$, K.P. Katin}

\vskip 4mm

\centerline{\it National Nuclear Research University MEPhI}
\centerline{\it Kashirskoe sh. 31, Moscow, 115409 Russia}

\vskip 5mm

$^{1)}$e-mail: AIPodlivayev@mephi.ru

\vskip 5mm

\centerline{\bf Abstract}

The molecular dynamics and Monte Carlo studies of the thermal stability of  C$_{20}$, C$_{36}$, and C$_{60}$ fullerenes and the methane molecule are reported. It has been shown that the heat transfer between the atomic cluster and the external heat reservoir can either promote or prohibit the formation of defects in this cluster. The wide temperature and pressure ranges have been determined where the defect formation rate is essentially non Arrhenius. It has been shown that the lifetime of light clusters in molecules depends more strongly on the contact with the heat reservoir. A statistical model that is based on the kinetic equation and allows for an analytical solution has been developed to explain the results. Within this model, the generalized Arrhenius formula has been derived to predict the lifetime of the clusters in an arbitrary thermal contact with the environment.

\vskip 5mm


\vskip 8mm

The mean lifetime $\tau$ of  an  atomic  cluster  to  its decay or transition to another isomer is often
described by the Arrhenius law
\begin{equation}
\tau(T)=\tilde{\xi} e^{E_a/kT},\label{eq:1}
\end{equation}
where $k$ $-$is the Boltzmann constant, $E_a$ $-$ is the decay
activation (isomerization) energy, $T$ $-$ is the temperature
of the environment in contact with the cluster, and the preexponential factor $\tilde{\xi}$ has the dimension of time and is the inverse of the frequency factor of the corresponding process. In some approaches, this preexponential factor is considered constant (see, e.g.,~\cite{bib:1}), but in the general case it depends not only on the atomic cluster structure and the characteristic heat transfer time
 $\tau_{th}$ between the cluster and environment, but also on the temperature, i.e.,
 $\tilde{\xi}=\tilde{\xi}(\tau_{th},T)$ (see, e.g.,~\cite{bib:2,bib:3}).  

If there is no thermal contact between the cluster and the environment 
 $(\tau_{th}\to+\infty)$, the temperature dependence of $\tau$ has another fundamentally nonArrhenius form~\cite{bib:4,bib:5}:  
\begin{equation}
\tau_0(E)=\left\{ \begin{array}{ll}
\xi\left(1-\frac{E_a}{E}\right)^{1-X},&\textrm{if }E>E_a\\
+\infty, &\textrm{if }E\leq E_a
\end{array}\right.
\label{eq:2}
\end{equation}
Here, $E=CT_m$, $E=E_{kin}+E_{pot}$ $-$ is the sum of the kinetic and potential energies of the cluster (the latter energy is measured from the energy of the equilibrium state of the cluster), where
 $C$ $-$ is the heat capacity of the cluster and $T_m$ $-$ is the microcanonical temperature~\cite{bib:6,bib:7}, relating to the average kinetic energy of the cluster
atoms $\langle E_{kin}\rangle$ as $\langle E_{kin}\rangle=\frac1{2}kT_mX$, where $X=(3N_{at}-6)$ $-$ is the number of degrees of freedom of the cluster consisting of $N_{at}$ atoms taking into account the conservation of three components of the total momentum and the angular momentum.

As the heat transfer intensity between the cluster and environment increases, lifetime dependence 
~(\ref{eq:2}), is transformed into Eq. ~(\ref{eq:1}). The key objective of this work is to study the character of the ~(\ref{eq:2}) to~(\ref{eq:1}) transformation, as the heat transfer time
 $\tau_{th}$ decreases to a finite value, and to determine the relation between the
preexponential factors $\tilde{\xi}$ and $\xi$ of the decay of the thermalized and heatinsulated clusters.

In a computer simulation, the quantity $\tau$ is determined as the average of the lifetimes of several clusters decayed at the same temperature (the energy for the heatinsulated clusters) but at different initial distributions of the atomic velocities in these clusters. If the number of clusters $M_0$ is sufficiently large, then the number of clusters $M(t)$, retained by the time $t$, decreases exponentially as  $M(t)=M_0exp(-t/\tau)$. At a sufficiently large number of statistical tests, the probability density $n$ finding the cluster in a certain state can be described in the framework of the kinetic equation~\cite{bib:8}. 

For the cluster of $N_{at}$ atoms, the $n$ value is generally a function of $6N_{at}$ coordinates of the phase space and the time~\cite{bib:8}. To solve the main problem of this work, it suffices to assume that $n$ depends only on the time $t$ and the total configuration energy $E$, i.e., $M(t)=\int \limits_{0}^{+\infty}n(E,t)dE$. Deriving the kinetic equation, it is covenient to consider that the clusters are introduced into the system continuously at a constant rate $I$ rather than simultaneously at the time $t=0$. In this case, the
initial energy of the cluster is $E_0=CT$ (or $CT_m$ for the heatinsulated cluster). It is also assumed that the cluster lifetime is much longer than the characteristic period of its natural oscillations and the characteristic time of the internal heat transfer between its normal modes. The kinetic equation of the particle balance is
\begin{equation}
\frac{\partial {n(E,t)}}{\partial t}=I\delta(E-E_0)-\frac{1}{\tau_0(E)}n(E,t)
+\frac{1}{\tau_{th}}\left(-n(E,t)+S(E,T)\int \limits_{0}^{+\infty}n(E',t)dE'\right),\label{eq:3}
\end{equation}
where $E_0=CT$, $S(E,T)$ $-$ is the equilibrium distribution density of an individual thermalized cluster in the energy $E$ at a temperature $T$. To avoid the consider ation of the heat transfer processes from the environ ment to the individual atoms of the cluster, we assumed that like in~\cite{bib:9}, the entire cluster is simultaneously thermalized; i.e., the nonequilibrium distribution of its states with the density $n(E,t)$ is transformed in the characteristic time interval $\tau_{th}$ to the equilibrium one with the density $S(E,T)$. 

The dependence of the average cluster lifetime $\tau$ on the characteristic thermaliza\-tion time $\tau_{th}$ and other parameters of interest can be obtained from the stationary solution $n_0(E)$ of Eq.~(\ref{eq:3}), 
\begin{equation}
0=I\delta(E-E_0)-\frac{1}{\tau_0(E)}n_0(E,t)+\frac{1}{\tau_{th}}\left(-n_0(E)+
S(E,T)\int \limits_{0}^{+\infty}n_0(E')dE'\right).\label{eq:4}
\end{equation}
For the steadystate process, the average lifetime $\tau$, the total number of the clusters $M(t)|_{t=+\infty}$ in the system, and the rate $I$ of their addition to the system are related as
\begin{equation}
\int \limits_{0}^{+\infty}n_0(E)dE=\tau I.\label{eq:5}
\end{equation}
Substituting Eq.~(\ref{eq:5}) into Eq.~(\ref{eq:4}), we obtain
\begin{equation}
n_0(E)=I\frac{S(E,T)\tau +\tau_{th}\delta(E-E_0)}{1+\tau_{th}/\tau _0(E)}.\label{eq:6}
\end{equation}
Integrating expression~(\ref{eq:6}) with respect to $E$ and taking into account Eq.~(\ref{eq:5}) and the relation $\int \limits_{0}^{+\infty}S(E,T)dE=1$, we obtain the desired dependence of $\tau$ on the parameters of the problem in the form
\begin{equation}
1/\tau=(1/\tau_{th}+1/\tau _0(CT))\int \limits_{0}^{+\infty}\frac{S(E,T)(1/\tau _0(E))}{1/\tau _{th}+1/\tau _0(E)}dE.\label{eq:7}
\end{equation}

The preexponential factor $\xi$ is independent of the activation energy but can depend on other parameters, such as the temperature~\cite{bib:3}, atomic mass, characteris\-tic thermalization time $\tau_{th}$, etc. In the frequent collision limit, $(1/\tau_{th}\to+\infty)$, the cluster decay can be considered as the diffusion of the atoms with a certain diffusion coefficient $D$ from the metastable equilibrium point through the energy barrier separating the cluster from its decay products~\cite{bib:2,bib:10}. In this case, the lifetime depends on $D$. Meanwhile, Eq.~(\ref{eq:7}) in the limit $1/\tau_{th}\to+\infty$ has the form
\begin{equation}
1/\tau=\int \limits_{0}^{+\infty}S(E,T)(1/\tau _0(E))dE.\label{eq:8}
\end{equation}
In this expression, the only parameter that can be a function of  $D$, is the preexpo- nential factor $\xi$ in Eq.~(\ref{eq:2}). For further analysis, it suffices to expand this parameter into the Taylor series up to the linear term in $1/D$. Taking into account that the diffusion coefficient for Brownian motion is proportional to the mean free path time and temperature~\cite{bib:8}, and the mean free path time is proportional to the thermalization time $\tau_{th}$, we have $D\sim\tau_{th}T$, and the preexponential factor $\xi$ can be represented as
\begin{equation}
\xi \approx\xi_0+\xi_1/(\tau_{th}T).\label{eq:9}
\end{equation}

We studied the transition of the heatinsulated decay mode to the thermalized one on an example of the four clusters of C$_{60}$, C$_{36}$, C$_{20}$ fullerene and CH$_4$ methane molecule. For the C$_{60}$ and C$_{36}$ fullerenes, we studied the isomerization process resulting from the Stone-Wales transformation~\cite{bib:11}. The decay process was conside\-red for the C$_{20}$ fullerene and CH$_4$ molecule. The  C$_{20}$ fullerene is the lightest of the existing fullerenes; for this reason, methane is added to the series of the clusters under study in order to follow the effect of the cluster mass on the character of the transition from law~(\ref{eq:2}) to law~(\ref{eq:1}). 

The fullerenes were simulated with the orthogonal tight-binding potential~\cite{bib:12}, successfully used in the simulation of both small clusters and macroscopic carbon structures (earlier, we applied it to investigate various characteristics of the C$_{60}$ and C$_{20}$ fullerenes and oter carbon systems~\cite{bib:13,bib:14,bib:15,bib:16,bib:17}). To determine the lifetime of the CH$_4$ molecule, we used the nonorthogonal
tightbinding method~\cite{bib:18}, which requires much more computer resources than the orthogonal tight-binding method, but was successfully used to determine the activation energy of the hydrocarbon structures ~\cite{bib:19}.

The parameters $E_a$, $\xi_0$ and $\xi_1$, in Eq.~(\ref{eq:7}), were determined by the numerical simulation of the decay or isomerization of the corresponding clusters. The activa\-tion energy $E_a$ of the processes of interest was determined by the static simulation method. Initially, the stationary points of the interaction potential in the ($3N_{at}-6$)-dimensional space of the generalized coordinates of the cluster atoms were found. These stationary points correspond to the metastable and saddle configura\-tions whose energies are used to determine the height of the energy barrier separating the initial configuration from the decay products (or the nearest isomer if the isomerization process is considered). The process activation energy was assumed as equal to this barrier height. The details of the static simulation method are presented in~\cite{bib:20}. In the general case, the activation energy can differ from the barrier height, but these two quantities for fullerenes coincide with each other~\cite{bib:21}. We also observed a similar pattern in methane.

We determined the cluster lifetime at different temperatures and heattransfer intensities by the molecular dynamics (MD) method and/or by the combined molecular dynamics and Monte Carlo (MDMC) method. The MD method was used for the temperatures at which the cluster lifetime did not exceed several nanoseconds. In this case, the Newton equation describing the motion of cluster atoms was numerically solved by the velocity Verlet method~\cite{bib:21}. The heat transfer with the environment was simulated by random collisions of the cluster atoms with the buffer gas atoms as in~\cite{bib:9}. In this work, we were interested in a wide range of the cluster thermalization times $\tau_{th}$ from $10^{-16}$\,s to $1$\,s. The fullerenes are synthesized at the buffer gas (helium or argon) pressure, which provides $\tau_{th}\approx 10^{-8}$\,s ~\cite{bib:Sid}, and the fast thermalization with the characteristic time $\tau_{th}=10^{-14}-10^{-16}$\,s can be obtained at the cluster thermal contact with the crystalline substrate. The MD method allows one to determine the lifetime (or the isomerization time) of the cluster at an arbitrary (including zero) intensity of the heat transfer between the cluster and the environment, but it requires large computer resources and is thus applicable only for sufficiently high temperatures at which the lifetime is short ($\sim1$\,ns).The MDMC method was applied to numerically determine the cluster lifetime at low temperatures at which the lifetime is $\sim1$\,s~\cite{bib:22}; this method is applicable at a sufficiently high heat transfer intensity when the $\tau_{th}$ value is comparable with the minimum period of the natural oscillations of a fullerene that is 24 and 20 fs for C$_{20}$ and C$_{60}$ respectively, in the tightbinding approximation. In their combination, the MD and MDMC methods allow one to investigate the entire ranges of temperatures and heattransfer intensities.

To compare the numerically simulated average cluster lifetimes to Eq.~(\ref{eq:7}), the cluster heat capacities $C$ and the distribution densities $S(E,T)$ were determined in the harmonic approximation
\begin{equation}
S(E,T)=\frac1{X!(kT)^{X+1}}E^Xe^{-E/kT}, C=kX.\label{eq:10}
\end{equation}
Approximation~(\ref{eq:10}) is justified by the data reported in~\cite{bib:23}, where it was shown that the energy distribution density in the C$_{20}$, fullerene calculated in the tight-binding model is almost identical to Eq.~(\ref{eq:10}).

The parameters $E_a$, $\xi_0$ and $\xi_1$ for the C$_{20}$, C$_{36}$ and C$_{60}$ fullerenes and the methane CH$_4$ molecule are presented in the table. The activation energy $E_a$ was determined by the static simulation method to a relative accuracy $\varepsilon \sim 10^{-4}$. The $\xi_0$ value was determined both by the comparison of Eq.~(\ref{eq:2}) and the lifetime for the heatinsulated clusters obtained by the MD method and by the comparison of Eq.~(\ref{eq:7}) and the lifetime for the thermalized clusters obtained by the MDMC and MD methods. The $\xi_1$ value was determined by comparing relation~(\ref{eq:7}) and the lifetime for the thermalized clusters obtained by the MDMC and MD methods. The number of numerical calculations $N_{var}$, determining the statistic error of $\xi_0$ and $\xi_1$, is also presented in the table.

\renewcommand{\tablename}{Table}
\begin{table}[h]

\caption{Activation energies $E_a$, preexponential factor $\tilde \xi$ and parameters $\xi _0$ and $\xi _1$ of the decay (isomerization) preexponential factor for the C$_{20}$, C$_{36}$, and C$_{60}$ fullerenes and methane molecule}
\begin{center}
\begin{tabular}{|c|c|c|c|c|}
\hline
 & $E_a$, eV & $\tilde \xi$, s & $\xi _0$, s & $\xi _1$, s$^2$K \\
 & Method & (1/$\tau_{th}$=0) & (1/$\tau_{th}\neq$0) & (1/$\tau_{th}\neq$0) \\
 &  & Method ($N_{var}$) & Method ($N_{var}$) & Method ($N_{var}$) \\
\hline
C$_{60}$ & 6.48 & 10$^{-17.31\pm 0.91}$ & 10$^{-17.17\pm 0.26}$ & 10$^{-36.13\pm 1.17}$ \\
  & SM & MD(40) & MDMC(16) & MDMC(16) \\
\hline
C$_{36}$ & 5.74 & 10$^{-15.72\pm 0.82}$ & 10$^{-15.58\pm 0.19}$ & 10$^{-35.33\pm 0.87}$ \\
  & SM & MD(40) & MDMC(24) & MDMC(24) \\
\hline
C$_{20}$ & 5.00 & 10$^{-17.60\pm 0.83}$ & 10$^{-17.46\pm 0.14}$ & 10$^{-35.21\pm 0.79}$ \\
  & SM & MD(40) & MDMC(24) & MDMC(24) \\
\hline
CH$_4$ & 3.55 & 10$^{-16.42\pm 0.88}$ & 10$^{-16.34\pm 0.02}$ & 10$^{-34.69\pm 0.43}$ \\
  & SM & MD(100) & MD(4000) & MD(4000) \\
\hline
\end{tabular}
\end{center}
\end{table}

The tabulated data imply that the $\tilde \xi$,value determined by the MD method in the heatinsulated clusters coincides to a statistical error with the $\xi_0$, value determined by the MDMC method for the clusters in contact with the heat reservoir. This indicates both the validity of the MDMC approach and the adequacy of analytical dependence~(\ref{eq:7}). The accuracy of the determination of $\xi_0$ by the MDMC method is higher than that for the MD method, because the MDMC method covers a wider temperature range.

We note that the dependences of the lifetimes of the methane molecule and fullerenes on the heattransfer intensity are different. Relation~(\ref{eq:7}) implies that the lifetime of the heatinsulated cluster at sufficiently low temperatures is infinite, $\tau(\tau_{th},T)|_{1/\tau_{th}=0}=\infty$ at $T<T_c$, $T_c=E_a/kX$ (since the cluster is heat insulated, the microcanonical temperature is assumed). For the
methane molecule and C$_{20}$, C$_{36}$, and C$_{60}$ fullerene, $T_c$= 4576\,K, 1074\,K, 653\,K and 432\,K, respectively. Therefo\-re, at the typical temperatures of petrochemical reactions $T\sim 800$\,K, the depen\-dence of the lifetime of the methane molecule on its heat transfer intensity is always nonmonotonic. Figure 1 presents this dependence at different temperatures. The figure implies that the methane molecule is least stable at the characteristic heat transfer time with the environment $\tau_{th}\sim10^{-13}-10^{-14}$\,s at any temperature. 

The heattransfer intensity dependence of the lifetime of comparatively heavy fullerenes has a different shape. Figure 2 presents the $\tau(\tau_{th},T)$ curves for the С$_{20}$ fullerene. The comparison of Figs. 1 and 2 shows that the decrease in the lifetime for the  С$_{20}$ fullerene with increasing heatexchange intensity is not so large, the minimum is more shallow and is similar to a plateau; i.e., the fullerene lifetime is independent of $\tau_{th}$ in a wide range of this parameter. For the C$_{36}$ and C$_{60}$ fullerenes, this trend gets more pronounced, because the clusters with a larger number of atoms are themselves heat reservoirs, and the defect formation processes in them depend much weaker on the interaction with the environment.

To conclude, we note that the dependence of the cluster lifetime on the heat transfer conditions presented in this work is more universal than the previous approximations. However, its applicability field is also limited by the assumption of the constant heat capacity of the cluster and its quasiharmonic behavior to the decay time. It would be interesting to extend this approach to the case of a molecule with pronounced
anharmonic oscillations.

We are grateful to L.A. Openov for useful discussions and critical comments.

\renewcommand{\refname}{\begin{center}{\Large\rm\bf REFERENCES}\end{center}}

\newpage
\includegraphics[width=15cm,height=15cm]{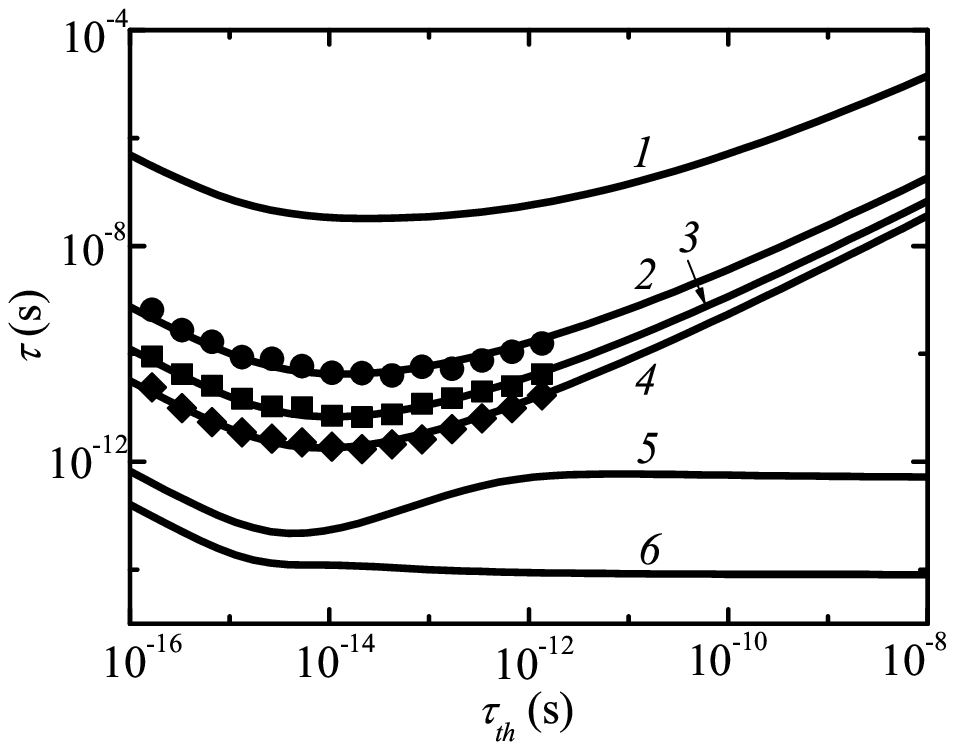}
\vskip 20mm
{\bf Fig.1.} Log-log plot of the lifetime $\tau$ of the methane molecule versus the charac- teristic heattransfer time $\tau_{th}$ with the thermostat at different temperatures. The curves are the theoretical dependence (7) for temperatures (from top to bottom) 10 000, 7000, 4000, 3500, 3000, and 2000 K. The circles, squares and rhombuses are the numerical calculation data for 4000, 3500, and 3000 K, respectively. Each symbol is the result of averaging over 100 independent calculations.

\newpage
\includegraphics[width=15cm,height=15cm]{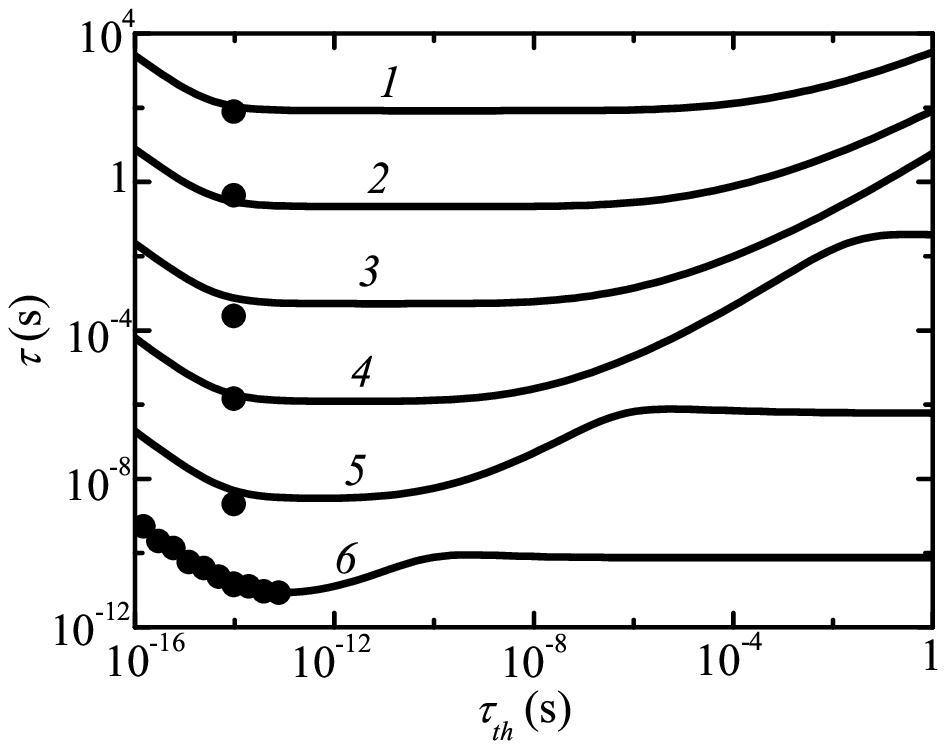}
\vskip 20mm
{\bf Fig.2.} Log-log plot of the lifetime $\tau$ of the C$_{20}$ fullerene versus the characteristic heattransfer time $\tau_{th}$ with the thermostat at different temperatures. The curves are the theoretical dependence (7) at temperatures (from top to bottom) 4000, 2824, 2182, 1777, 1500, and 1300 K. The circles are the numerical calculation data. Each circle is the result of averaging over 16 independent calculations.

\end{document}